\documentclass[sigconf,screen=true,anonymous=false,bookmarks=true,acm]{acmart}

\AtBeginDocument{%
  }
    
\usepackage{multirow}
\usepackage{subcaption}
\usepackage{bbding}
\usepackage{mathtools}
\usepackage{graphicx}
\usepackage{diagbox}
\usepackage{textcomp}
\usepackage{algorithm}
\usepackage{algpseudocode}
\algtext*{EndFor}
\algtext*{EndFunction}
\algtext*{Until}
\algtext*{EndIf}
\algtext*{EndWhile}
\usepackage{amsmath}
\usepackage{array}
\usepackage[referable]{threeparttablex}
\usepackage[T1]{fontenc}
\usepackage[utf8]{inputenc}
\usepackage{tablefootnote}
\usepackage{geometry}

\usepackage{tikz}

\usepackage{listings}
\usepackage{xcolor}

\definecolor{codegreen}{rgb}{0,0.6,0}
\definecolor{codegray}{rgb}{0.5,0.5,0.5}
\definecolor{codepurple}{rgb}{0.58,0,0.82}
\definecolor{backcolour}{rgb}{0.95,0.95,0.92}

\lstdefinestyle{mystyle}{
  backgroundcolor=\color{backcolour},   commentstyle=\color{codegreen},
  keywordstyle=\color{magenta},
  numberstyle=\tiny\color{codegray},
  stringstyle=\color{codepurple},
  basicstyle=\ttfamily\footnotesize,
  breakatwhitespace=false,         
  breaklines=true,                 
  captionpos=b,                    
  keepspaces=true,                 
  numbers=left,                    
  numbersep=5pt,                  
  showspaces=false,                
  showstringspaces=false,
  showtabs=false,                  
  tabsize=2
}
\usepackage{colortbl}
\usepackage{dcolumn}

\definecolor{greenDeep}{RGB}{0,170,0}
\definecolor{greenSlightDeep}{RGB}{0,205,0}
\definecolor{greenShallow}{RGB}{0,255,0}
\definecolor{greenShallower}{RGB}{160,255,0}
\definecolor{orangeShallow}{RGB}{255,190,0}
\definecolor{orangeDeep}{RGB}{255,80,0}
\definecolor{orangeDeeper}{RGB}{255,40,0}
\definecolor{redDeep}{RGB}{255,0,0}

\definecolor{redLight}{RGB}{255,128,114}

\def\zz#1{%
\ifdim#1pt>4.9pt\cellcolor{greenDeep}\else
\ifdim#1pt>3.9pt\cellcolor{greenSlightDeep}\else
\ifdim#1pt>2.9pt\cellcolor{greenShallower}\else
\ifdim#1pt>2.9pt\cellcolor{yellow}\else
\ifdim#1pt>1.9pt\cellcolor{orangeShallow}\else
\ifdim#1pt>1.9pt\cellcolor{orange}\else
\ifdim#1pt>0.9pt\cellcolor{orange}\else
\ifdim#1pt>0.9pt\cellcolor{orangeDeep}\else
\cellcolor{orangeDeep}\fi\fi\fi\fi\fi\fi\fi\fi
#1}

\graphicspath{{./_fig/}}

\makeatletter
\renewcommand\footnoterule{%
  \kern-3\p@
  \hrule\@width0.4\columnwidth
  \kern2.6\p@}
  \makeatother

\definecolor{lightgreen}{RGB}{198, 224, 183}
\definecolor{lightred}{RGB}{240, 205, 176}

\usepackage{pifont}

\usepackage[hang,flushmargin]{footmisc}

\copyrightyear{2026}
\acmYear{2026}
\setcopyright{cc}
\setcctype{by}
\acmConference[DAC '26]{63rd ACM/IEEE Design Automation Conference}{July 26--29, 2026}{Long Beach, CA, USA}
\acmBooktitle{63rd ACM/IEEE Design Automation Conference (DAC '26), July 26--29, 2026, Long Beach, CA, USA}
\acmDOI{10.1145/3770743.3804229}
\acmISBN{979-8-4007-2254-7/2026/07}

\begin{document}

\title{COOL: A Cooling-Aware Point Transformer Framework for Thermal Prediction in Advanced 3D/3.5D IC Packaging}
\author[1]{Yao Lu, Zhicheng Guo, Qijun Zhang, Shang Liu, Wenji Fang, Wenkai Li, Zhiyao Xie$^*$\\ 
Hong Kong University of Science and Technology\\
Email: eezhiyao@ust.hk \\
($^*$corresponding author)}

\begin{abstract}


Advanced 3D and 3.5D IC packaging significantly improves integration density but elevates thermal management challenges due to cross-layer heat coupling and complex cooling structures. Traditional solvers deliver high fidelity but are too slow for iterative design flows, while existing learning-based methods either fail to capture inter-die thermal coupling or treat cooling structures as static components, limiting their applicability in real packaging co-design scenarios.
In this work, we introduce COOL, a cooling-aware point transformer framework that represents heterogeneous assemblies (dies, interposers, TIMs, heat spreaders) as annotated 3D point clouds embedding geometric, material and power attributes. COOL explicitly encodes geometric boundaries and cooling structures, and introduces a physics-informed boundary condition (PI-BC) loss to enforce thermal consistency at material interfaces and cooling boundaries. Extensive experiments demonstrate that COOL achieves a remarkable 2.4\% NMAE on our constructed benchmark of multi-package thermal designs, substantially outperforming existing learning-based approaches while providing over 15.7× speedup compared to commercial FEM solvers. 
   
\end{abstract}

\maketitle
\pagestyle{plain}

\section{Introduction}
\label{sec:introduction}

Advanced 3D and 3.5D IC packaging technologies enable high integration density and performance, but also introduce severe thermal coupling and non-uniform heat dissipation challenges. Accurate thermal analysis is essential for ensuring performance and reliability. However, conventional finite-element method (FEM) or finite-difference method (FDM) solvers such as COMSOL, Cadence Celsius, and ANSYS Icepak are computationally expensive, often requiring hours to simulate a single power trace. In addition, these solvers require substantial manual effort for geometric modeling and meshing. This makes them impractical for iterative design space exploration or layout optimization automatically.

\begin{figure}[!t]
\centering
\includegraphics[width=0.48\textwidth]{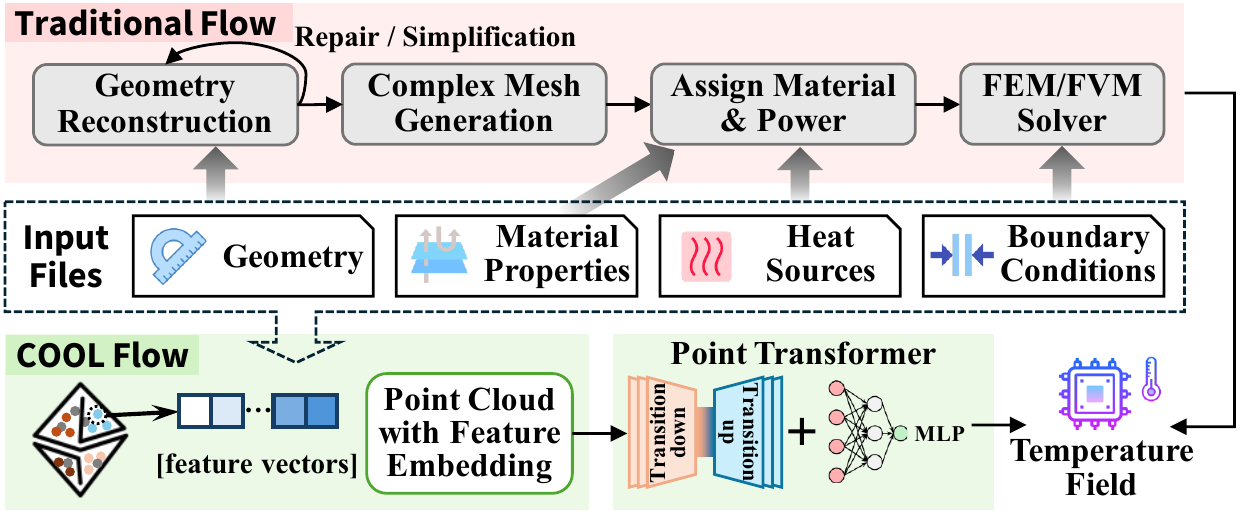}
\vspace{-0.2in}
\caption{Comparison between the traditional thermal workflow and the proposed COOL framework. Traditional FEM/FVM workflows require manual preprocessing for geometry, meshing, and boundary setup. COOL encodes all inputs into a unified point-cloud embedding and uses a point transformer for direct temperature prediction, enabling fast and automated thermal analysis.
}
\label{fig:intro}
\vspace{-.4in}
\end{figure}

In realistic 3D/3.5D packaging, multiple active dies are vertically stacked and thermally coupled through inter-die interfaces~\cite{wang2024exploiting,pak2014electromigration} such as thermal interface materials (TIMs), through-silicon vias (TSVs), and micro-bumps.  
At the same time, chiplets often differ in thickness and material composition, resulting in heterogeneous and highly anisotropic heat transfer paths.  
These cross-layer interactions and geometric variations jointly determine the temperature field within the package, making accurate modeling far more complex than in single-die systems~\cite{feng2022chiplet,shao2019simba,chen2020fast}.  
Recent learning-based thermal models~\cite{huang2025self, zhou2025asrr, wang2025hisim, chen2022fast, liu2023deepoheat, yu2025deepoheat, chen2023fast} have achieved promising speedup over traditional FEM solvers. However, they are inadequate for capturing complex thermal dynamics in advanced packaging. These approaches face serious limitations in modeling inter-die thermal coupling and adapting to design-dependent cooling paths, restricting their applicability to real 3D/3.5D IC scenarios as follows.

First, existing models \textbf{fail to capture inter-die thermal coupling across multiple stacked layers}.  
For example, DeepOheat~\cite{liu2023deepoheat} and DeepOheat-v1~\cite{yu2025deepoheat} learn temperature distributions from 2D or volumetric power maps only within a single active die, their formulations do not model heat propagation between vertically stacked chiplets.
Similarly, ASRR-PINN~\cite{zhou2025asrr} assumes a uniform power map at one cross-section and does not vary power across layers, over-simplifying treating the system as a single-die structure.
Such simplifications fail to capture the vertical conduction and lateral spreading effects that dominate real 3D IC thermal behavior.

Second, current learning frameworks \textbf{treat cooling structures as static fixed conditions rather than flexible design elements with variation}.  
Both DeepOHeat models~\cite{liu2023deepoheat, yu2025deepoheat} and ASRR-PINN~\cite{zhou2025asrr} assume fixed boundary conditions, ignoring the influence of cooling structures such as heat spreaders or TIMs. SAU-FNO~\cite{huang2025self} makes progress by explicitly modeling boundary constraints and include cooling-related layers like heat sinks or TIMs. However, its formulation treats these components as pre-defined, hard-coded layers with fixed material properties and thickness.  
Such rigid modeling cannot capture the design-dependent variations commonly found in advanced packaging, where TIM thickness and chiplet stacking geometry may all be adjusted dynamically during co-design. 
As a result, existing approaches fail to generalize across packaging configurations with heterogeneous structural variability.

To overcome these challenges, a more flexible geometric representation is needed to model the complex, heterogeneous structures of 3D/3.5D packages, along with a corresponding model capable of rapidly capturing information from this new data format. 



In this work, we propose \textbf{COOL}, the first cooling-aware data-driven framework that jointly learns heat transfer across chips, packages, and cooling components in Figure~\ref{fig:intro}. We propose two key techniques: \textbf{data format} and \textbf{point transformer model}.
\ding{182} COOL represents the heterogeneous advanced packaging structures by point cloud, which can naturally encode spatial and material heterogeneity at arbitrary resolutions. Each point can represent a physical location with attributes such as material type, power distribution, or boundary conditions, enabling modeling of thermal behavior across irregular geometries and die interfaces. 
\ding{183} COOL applies transformer-based attention mechanisms on this point-cloud formulation as the foundation of our thermal learning framework. To incorporate more physical information, we customize COOL to be aware of the boundary conditions. We use the point-based neural operators~\cite{zhao2021point, wu2024point}, which demonstrate strong capability in learning spatially continuous physical fields directly from point sets, achieving both geometric flexibility and data efficiency.

In summary, by representing the 3D thermal domain as a point cloud and applying transformer-based attention mechanisms, COOL captures complex geometric and material interactions efficiently. 
This unified formulation enables efficient and generalizable temperature prediction across diverse 3D packaging configurations. 
\textbf{Our key contributions are summarized as follows:}
\vspace{-0.1in}
\begin{itemize}
    \item \textbf{COOL} is the first learning-based thermal prediction framework that explicitly models the coupled heat transfer among chips, packages, and cooling structures in realistic 3D and 3.5D IC packaging.  
    This formulation captures the design-dependent effects of TIMs, TSVs, and heat spreaders, which are overlooked by existing grid-based or single-die approaches. 
    
    \item We propose a \textbf{point-cloud-based learning methodology} that represents the entire thermal domain as a collection of spatial points, each annotated with its material property, structural domain type, and power activity.  
    
    \item We introduce \textbf{physics–informed boundary conditioning (PI–BC)} in our learning framework, which injects boundary physics into training, enforcing boundary-conditioned heat transfer and inter-die coupling to enhance accuracy at material interfaces and hotspot regions.
    
    \item We construct diverse thermal datasets covering multiple 3D/3.5D packaging configurations. 
    Comprehensive experiments demonstrate that COOL achieves fast convergence and a speedup of 15.7$\times$ over commercial FEM solvers, with only an NMAE of around 2.4\%, which is significantly lower than the baselines on the testing set. 
    
\end{itemize}

         

\section{Background}
\label{sec:background}

\begin{figure}[!t]
\centering
\vspace{-.1in}
\includegraphics[width=0.45\textwidth]{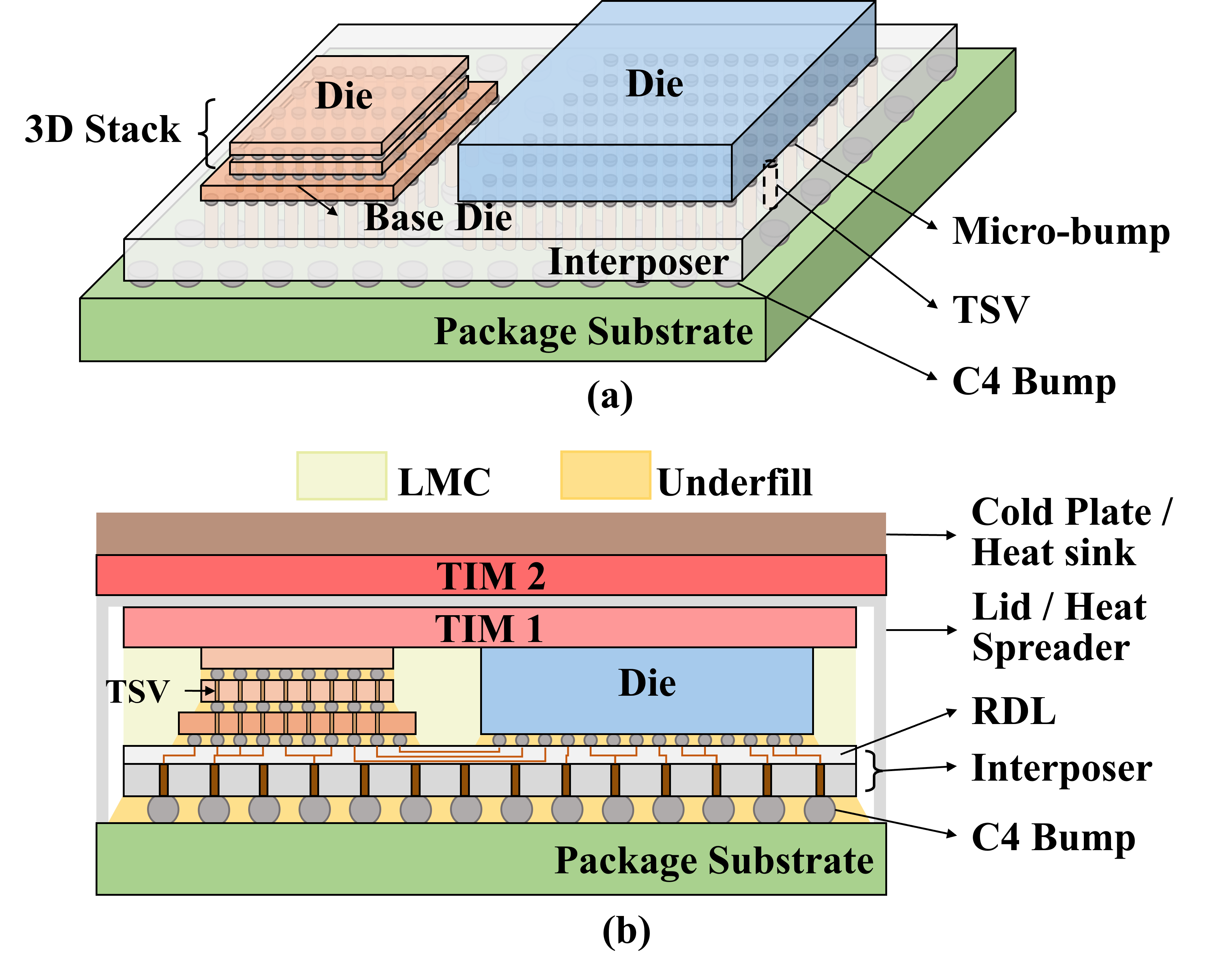}
\vspace{-.2in}
\caption{The structure of 3.5D integrated chips, a fusion of 2.5D and 3D-ICs. (a) is a 3D schematic view showing the heterogeneous integration of multiple chiplets on a silicon interposer. (b) Cross-sectional view illustrating CoWoS-based packaging with thermal management structures, including heat spreader, TIM, and heat dissipation devices.}
\label{fig:3dic}
\vspace{-.25in}
\end{figure}

In this section, we review background on advanced IC packaging and learning-based thermal prediction. We first discuss modeling challenges in 2.5D, 3D, and 3.5D stacks, then summarize prior approaches for incorporating boundary conditions in thermal models.

\subsection{Advanced Packaging Modeling}
Advanced packaging technologies such as 2.5D and 3D integration have emerged as key enablers for continuing system scaling beyond the limits of traditional monolithic SoCs. In particular, \textbf{3.5D integration} combines the advantages of both 2.5D interposer-based and 3D stacked architectures, enabling heterogeneous integration of multiple high-performance chiplets on silicon interposers with vertical interconnects for high bandwidth and low latency. 

Because COOL is explicitly designed for these practical 3D/3.5D configurations, it is necessary to model the geometric and material complexity inherent to such packages. Figure~\ref{fig:3dic} illustrates a representative 3.5D structure, where chiplets are bonded onto a silicon interposer through micro-bumps, and heat transfer paths involve multiple materials such as silicon, underfill, die attach, TIMs, and heat spreaders.  
Accurately capturing these structures requires detailed 3D representations, yet meshing fine-pitch objects such as C4 bumps and TSV arrays is computationally prohibitive.  

To build reliable thermal models for learning, we adopt physically meaningful geometric abstractions that retain thermal behavior while remaining simulation-friendly.  
C4 bumps are approximated as \textbf{regularized cubic} primitives with equivalent thermal conductivity, ensuring consistent heat-flow representation without prohibitive meshing cost.  
Similarly, dense TSV arrays are handled through \textbf{area-normalized material homogenization}: instead of meshing thousands of individual TSVs, we compute an effective anisotropic conductivity based on TSV density within each chiplet region.  
These techniques allow us to simulate real 3.5D structures with high fidelity and produce temperature fields.

\begin{figure*}[!t]
\centering
\vspace{-.5in}
\includegraphics[width=0.96\textwidth]{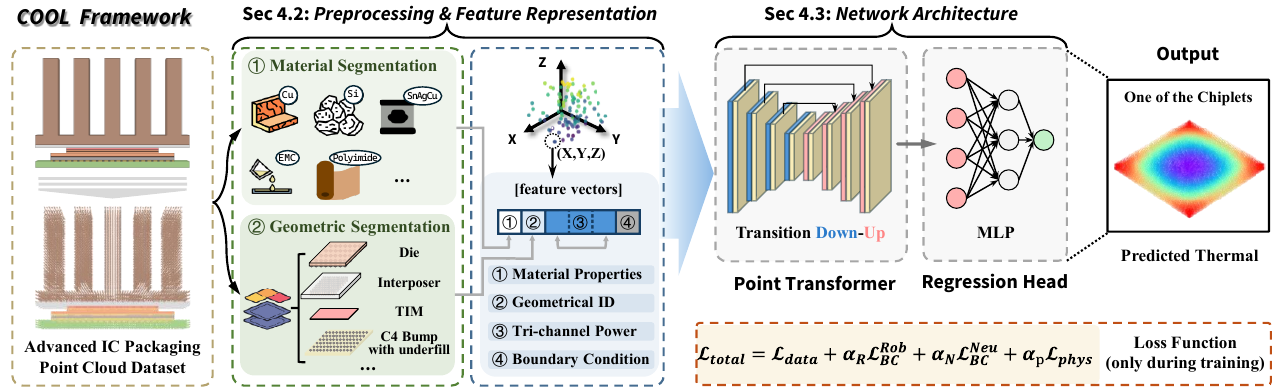}
\vspace{-.1in}
\caption{The framework of COOL. Point clouds representing heterogeneous 3D/3.5D IC assemblies are preprocessed with material and structural features and processed by a boundary-aware point transformer. 
A regression head predicts per-point steady-state temperatures, while the training loss enforces data fidelity, boundary conditions, and physics-inspired consistency.}
\label{fig:framework}
\vspace{-.15in}
\end{figure*}

\subsection{Learning-based Thermal Prediction} 
Learning-based thermal prediction has emerged as a promising approach for accelerating thermal analysis in complex IC stacks.  A central challenge is accurately incorporating \textbf{boundary conditions} (e.g., Robin, Neumann) and \textbf{material transitions}, which strongly influence heat spreading in 3D/3.5D packages.

Prior studies~\cite{Sun2025ChipletEMP2, chen2025neuralmesh, huang2025self, zhou2025asrr, wang2025hisim} have explored various mechanisms for embedding boundary information into neural models. For example, NeuralMesh~\cite{chen2025neuralmesh} enriches the network input by decomposing the 3D structure into material-transition layers and appending per-plane tensors that encode material properties, heat sources, and boundary attributes. ASRR-PINN~\cite{zhou2025asrr} 
enforces analytic satisfaction of 3D thermal constraints through a dedicated boundary-conditioning module.  
Despite these efforts, enforcing boundary physics in highly heterogeneous IC stacks remains challenging due to complex material interfaces and heat-dissipation paths. 

These limitations motivate our design of COOL. 
We adopt a \textbf{hierarchical point-based transformer architecture} that directly processes 3D point clouds of advanced packages, avoiding grid quantization error.  
To ensure physical correctness, COOL integrates a scalable \textbf{physics-informed boundary condition (PI-BC)} scheme that incorporates Robin and Neumann interactions directly into the training objective.  
This combination yields a cooling-aware framework suited for realistic 3D/3.5D IC thermal prediction.



\section{Problem Formulation}

Each 3D/3.5D IC design is represented as a point set and the task is to learn a mapping that outputs the steady-state temperature.
\begin{equation}
P=\{(\mathbf{x}_i,\mathbf{f}_i)\}_{i=1}^{N}\rightarrow (\hat{T}_1,\dots,\hat{T}_N)
\end{equation}

where $N$ is the number of the points in a design, $\hat{T}_i$ is the temperature of the point $i$, $\mathbf{x}_i\in R^3$ are the coordinates and $\mathbf{f}_i$ contains material-domain features, engineered power descriptors, and geometric/boundary indicators as described in Sec.~\ref{subsec:preprocessing}.  

This defines the point-based thermal prediction problem addressed by COOL: predicting physically consistent temperature fields directly from a unified point-cloud embedding.

\section{COOL Framework} \label{sec:COOL_framework}

This section introduces the architecture and design of the proposed COOL framework as illustrated in Figure~\ref{fig:framework}. Since we are the first to formulate thermal prediction for advanced IC packaging using point clouds, we develop a dedicated data preprocessing (in Section~\ref{subsec:preprocessing}) and a corresponding point transformer (in Section~\ref{subsec:architecture}) backbone that unifies chip and cooling structures. 

\subsection{Overview}
\label{subsec:overview}
The proposed COOL is a \textbf{point transformer framework} for steady-state temperature prediction in heterogeneous 3D/3.5D IC packages. 
Unlike traditional FEM solvers that discretize volumetric geometries onto fixed meshes, COOL directly operates on \textit{point-level representations}, maintaining both spatial fidelity and scalability. 


The framework is designed to (1) efficiently scale to designs containing hundreds of thousands of spatial points, (2) explicitly encode heterogeneous material domains and cooling structures, and (3) enhance the learnability of boundary-aware thermal distributions through carefully engineered representations. 

\subsection{Preprocessing and Feature Representation}
\label{subsec:preprocessing}
The COOL framework begins by transforming each 3D/3.5D IC design into a unified point-based thermal representation.
On average, a single design contains around $3\times10^5$ points, which capture fine-grained structural and material variations across multiple stacked dies, interconnect layers, and packaging materials. This fine-resolution representation enables COOL to model cross-die thermal coupling and heat spreading paths with high fidelity.

Coordinates $(X, Y, Z)$ are normalized to the interval $[0,1]^3$.  
Each point includes intrinsic features such as intrinsic thermal-related properties, geometric boundaries, and power dissipation values.

\subsubsection{Material-domain Features.}
\label{subsubsec:material}
Thermal conduction in advanced packaging is governed by interactions among highly heterogeneous materials, such as silicon, copper, underfill, and polymer. 
However, these materials often exhibit overlapping physical ranges of thermal conductivity ($k_{iso}$), specific heat ($C_p$) and density ($\rho$), which makes it difficult for purely continuous representations to distinguish domains unambiguously. 

To address this, COOL adopts a dual-domain encoding scheme. 
Each material domain is first assigned a categorical identifier that is one-hot encoded and concatenated with its continuous attributes $\{C_p, \rho, k_{iso}\}$. 
The combined vector is then projected into a material embedding space, enabling the model to learn both domain-specific conduction patterns and cross-material transition behavior. This can help COOL capture heat transfer across different materials. However, many previous works~\cite{chen2025neuralmesh, zhou2025asrr} fail to capture such specific material information in their learning model. 

\subsubsection{Comprehensive Power Features.}
\label{subsubsec:power}
Power dissipation ($q$), serving as the heat source, is extremely sparse and exhibits strong spatial skewness in real 3D IC systems. 
To capture this distributional complexity, COOL employs an engineered tri-channel power representation:
\textbf{1) Raw Power:} retains the absolute magnitude of dissipation, preserving direct correspondence to thermal intensity.
\textbf{2) Binary Flag:} indicates active versus passive regions. It helps guide the model to focus attention on power-generating sites.
\textbf{3) Log-Transformed Power:} converts long-tailed power distributions. It helps stabilize gradient flow during training.

These comprehensive representations enable COOL to specialize on different source regions, enhancing the model’s awareness of local heat density and spatial coherence.

\subsubsection{Geometric and Boundary Features}
Accurate steady-state thermal prediction strongly depends on how well the model captures geometric complexity and boundary-driven heat transfer behavior. 
To handle this, COOL augments each point with a set of geometry-aware encodings. First, different geometry domains are one-hot encoded. Then, boundary-type indicators are assigned to points whose coordinates lie on heat-exchange surfaces, including (1) convection boundaries flag ($b^{\text{conv}}_i$) and (2) adiabatic boundaries flag ($b^{\text{neu}}_i$). 
These features help the network localize boundary conditions that govern heat entering or exiting the system.

Overall, these encoding strategies yield a unified and physics-aware point representation for 3D/3.5D ICs. By embedding material properties, power descriptors, and geometry–boundary cues into each point, COOL provides the essential physical context required by the downstream transformer.


\subsection{Network Architecture}
\label{subsec:architecture}
COOL adopts a hierarchical point-based transformer architecture. We extend it with thermal-aware sampling and physics-informed constraints to better capture thermal behavior, cross-material conduction, and boundary-driven heat exchange that are critical in 3D/3.5D ICs. 
The network follows the encoder–decoder paradigm commonly used in geometric deep learning~\cite{Bronstein_2017}. 

\subsubsection{Encoder with Transition Down}
The encoder progressively reduces point resolution with several transition down (TD) layers and point transformer blocks while increasing feature capacity, enabling the network to reason jointly about local conduction patterns and global heat-spreading structures.

\ding{192} \textbf{Transition Down}. Each TD stage consists of sampling, neighborhood grouping, feature aggregation.

\emph{Sampling.}  
Standard farthest point sampling (FPS) relies solely on geometry and may overlook thermally important regions such as hotspots and interfaces.  
COOL incorporates \textbf{temperature–aware farthest point sampling (TA–FPS)}, which biases FPS using an importance score that reflects local thermal activity and material transitions.  
To quantify importance, TA–FPS computes a composite score $s_i \in [0,1]$:
\begin{equation}
s_i = \alpha_1 \hat{g}_i 
    + \alpha_2 \hat{q}_i 
    + \alpha_3 \hat{b}_i 
\end{equation}

$\hat{g}_i$, $\hat{q}_i$, and $\hat{b}_i$ are normalized temperature gradient magnitude, normalized power density, and boundary indicator, respectively.  
The classical FPS distance metric $D_i$ is biased as $D'_i = (1 + \gamma s_i)\, D_i$.

This sampling strategy ensures that regions influencing heat flow are retained at coarser resolutions, improving the fidelity of cross-scale thermal reasoning.

\emph{Neighborhood Grouping.}  
For each sampled point, a local neighborhood is constructed using $k$-nearest neighbors ($k$-NN)~\cite{qi2017pointnetdeephierarchicalfeature}. Relative coordinates and local feature differences are concatenated to provide spatial and thermal context for aggregation.

\ding{193} \textbf{Point Transformer Block}. Each TD stage includes a point transformer module that refines local representations using attention mechanisms.  
Given a point’s latent vector and its surrounding neighbors, the module computes attention weights based on learned positional encodings and feature affinity.  This allows the network to adaptively prioritize neighbors that influence heat transfer.

\subsubsection{Decoder with Transition Up}
The decoder mirrors the encoder and gradually restores the prediction to full resolution.

\textbf{Transition Up (TU)}.
Each TU stage upsamples features from a lower-resolution point set to a higher-resolution one.

\emph{Interpolation.}  
Features are propagated using distance-weighted interpolation over a small $k$-NN neighborhood, ensuring smooth reconstruction across materials and interfaces.

\emph{Skip Connections.}  
Encoder features are aligned and concatenated with decoder features to restore fine-grained structure lost during downsampling.

\subsubsection{Global Structure and Regression Head}
Across the hierarchy, the feature dimension expands from $d=128$ at the input embedding to progressively higher capacities at deeper layers.  
At the end of the decoder, a two-layer MLP maps point features to a scalar temperature prediction.

\subsubsection{Physics–Informed Boundary Conditioning}
Thermal behavior in packaged ICs is strongly influenced by boundary interactions, including heat spreading through heat sinks, convection at exposed surfaces, and continuity of flux across material interfaces.  

To capture these effects, COOL introduces PI-BC, which operates jointly with the data supervision loss.
The total loss is defined as
\begin{equation}
L = L_{\text{data}} 
  + \alpha_{\text{R}} L_{\text{BC}}^{Rob}
  + \alpha_{\text{N}} L_{\text{BC}}^{Neu}
  + \alpha_{\text{p}} L_{\text{phys}}
\end{equation}
where $\alpha_{\text{R}}$, $\alpha_{\text{N}}$, and $\alpha_{\text{p}}$ are coefficients.

\textbf{Data term:}  
$L_{\text{data}}$ is the mean squared error of $N$ pints between predicted $\hat{T}$ and ground-truth $T$ temperatures.
\begin{equation}
    L_{\text{data}}=\frac{1}{N}\sum_{i=1}^{N} \left( \hat{T}_i - T_i \right)^2 .
\end{equation}

\textbf{Physics term:}  
Temperature within a homogeneous material domain should remain spatially smooth except near heat sources.  
To discourage non-physical jumps, we introduce $L_{\text{phys}}$, a Laplacian-based regularization that penalizes deviations between each point and a weighted average of its same-material neighbors.

\textbf{Boundary condition term:} 
The boundary loss terms $L_{\text{BC}}^{Rob}$ and $L_{\text{BC}}^{Neu}$ are formulated similarly to the data term, but are selectively applied to points belonging to Robin and Neumann boundaries~\cite{pfaff2021learningmeshbasedsimulationgraph}, respectively. For each boundary type, we compute the mean squared error between predicted and ground-truth temperatures.

\section{Experiments}


\subsection{Experimental Setup}
\label{sec:setup}
\subsubsection{Dataset Construction.}




We first generate a diverse dataset of $74$ heterogeneous 3D and 3.5D IC packaging designs. 
These designs are constructed by systematically varying key structural parameters listed in Table~\ref{tab:para}. 
The parameter set spans multiple aspects of advanced packaging, including:  
(1) \textbf{chip and substrate geometry} (e.g., lateral dimensions $L_{x,y}$, die and interposer sizes) that determine heat–spreading distances;  
(2) \textbf{vertical stacking configurations} such as the number of dies $N_{\text{die}}$ and number of stacks $N_{\text{stacks}}$, covering both monolithic dies and multi-die 3D stacks;  
(3) \textbf{interconnect-related thermal paths}, including TSV density $f_{\text{TSV}}$ and C4 bump parameters, which influence vertical heat conduction;  
and (4) \textbf{cooling-structure parameters} such as TIM, lid, and heat-sink thicknesses $(L_z^{\text{TIM}}, L_z^{\text{lid}}, L_z^{\text{hs}})$ directly affecting heat extraction efficiency.  
Sampling each variable within the ranges of Table~\ref{tab:para} yields design instances that collectively various 3D/3.5D designs.

For each of the 74 generated designs, we additionally create 10 power-map variants, resulting in a total of \textbf{740} simulated thermal scenarios. 
Steady-state thermal fields are then computed using an FEM solver under realistic boundary and cooling conditions. 
The resulting temperature distributions are sampled into 3D point clouds, forming the dataset used for training and evaluating \textbf{COOL}.

To avoid geometric leakage, splits are performed at the design level. 
We randomly choose 80\% of the designs for training and 20\% for testing. Based on the different structures of the designs, there are a total of 58 designs, generating 580 training samples. The remaining 16 designs, with 160 samples, are used for testing. 

\subsubsection{Preprocessing}
To enable batch training and ensure architectural consistency, the original point clouds are resampled to a fixed size of $300{,}000$ points using farthest point sampling (FPS).
For each point, we construct the feature vector defined in Section~\ref{subsec:preprocessing}.  

The process includes:  
(1) normalized 3D coordinates (\textbf{3} dimensions);  
(2) material-domain descriptors, consisting of an \textbf{7}-dimensional one-hot material ID and \textbf{3} physical properties ($k_{iso}$, $C_p$, $\rho$);  
(3) \textbf{11}-dimensional geometry one-hot encoding and \textbf{2}-dimensional boundary indicators, and power-related features.  

Concatenating all components yields a 26-dimensional raw feature vector.   
Before entering the network, this vector is mapped to a 64-dimensional through a feature embedding layer.

\subsubsection{Model Setting}
 
COOL follows the hierarchical encoder–decoder architecture in Section~\ref{subsec:architecture} with fixed-resolution point inputs.  
The encoder performs three transition down stages, reducing the point set from 
$300\text{K} \rightarrow 150\text{K} \rightarrow 75\text{K} \rightarrow 37.5\text{K}$, 
while expanding feature channels as 
$64 \rightarrow 128 \rightarrow 256 \rightarrow 512$.  
Each stage includes TA–FPS sampling and a point transformer block.  
The decoder mirrors this hierarchy with symmetric Transition Up stages using interpolation-based feature propagation and skip connections.  

A lightweight MLP head ($64 \rightarrow 64 \rightarrow 1$) produces normalized temperature predictions for all points.

\begin{table}[!t]
\centering
\renewcommand{\arraystretch}{1.1}
\resizebox{0.42\textwidth}{!}{
\begin{tabular}{c|c|c}
\toprule
\textbf{Variable} & \textbf{Description} & \textbf{Range [Unit]} \\
\midrule\midrule

$L_{x,y}$ & Substrate lateral size (x-y) & 8-16 [mm] \\
$L_z$ & Substrate thickness & 400 [µm] \\

\midrule
$L_x^{\text{Si}}$ & Interposer lateral size & 6-14 [mm] \\

$L_{\text{C4}}$ & C4 bump height & 80-120 [µm] \\

$f_{\text{TSV}}$ & TSV density in interposer & 5--30\% of Si area \\
\midrule
$N_{\text{stacks}}$ & Number of stacks & {1,2,3} \\
$N_{\text{die}}$ & Number of stacked dies & {2,3,4} \\

$L_{x,y}^{(stack)}$ & 3D-stacked IC lateral size & 4-7 [mm] \\
$L_z^{(stack)}$ & 3D-stacked IC thickness & 100-250 [µm] \\

$L_{x,y}^{(mono)}$ & Monolithic die lateral size & 6-10 [mm] \\
$L_z^{(mono)}$ & Monolithic die thickness & 300-800 [µm] \\
\midrule

$L_z^{\text{TIM}}$ & TIM thickness & 100-220 [µm] \\
$L_z^{\text{lid}}$ & Lid thickness & 500-800 [µm] \\
$L_z^{\text{hs}}$ & Heat sink thickness & 600-1800 [µm] \\

\bottomrule
\end{tabular}
}
\caption{Design variables and parameter ranges used for constructing the 3D/3.5D IC thermal dataset. 
}
\label{tab:para}
\vspace{-.3in}
\end{table}

\subsubsection{Training and Testing}
Training uses the AdamW optimizer with initial learning rate $1\times 10^{-3}$ and weight decay $1\times10^{-4}$.  
A ten-epoch linear warm-up is followed by cosine annealing.  
Distributed data-parallel training on eight NVIDIA RTX 4090 GPUs supports batch size = 16.  
The full training schedule runs for 200 epochs and requires approximately 35.6 hours of wall-clock time.  
During testing, inference is executed on a single RTX 4090 GPU. The predicted normalized temperatures are inverse-transformed to obtain physical temperature fields. 


\begin{figure}[!b]
\centering
\begin{subfigure}{0.49\linewidth}
  \centering
  \includegraphics[width=\linewidth]{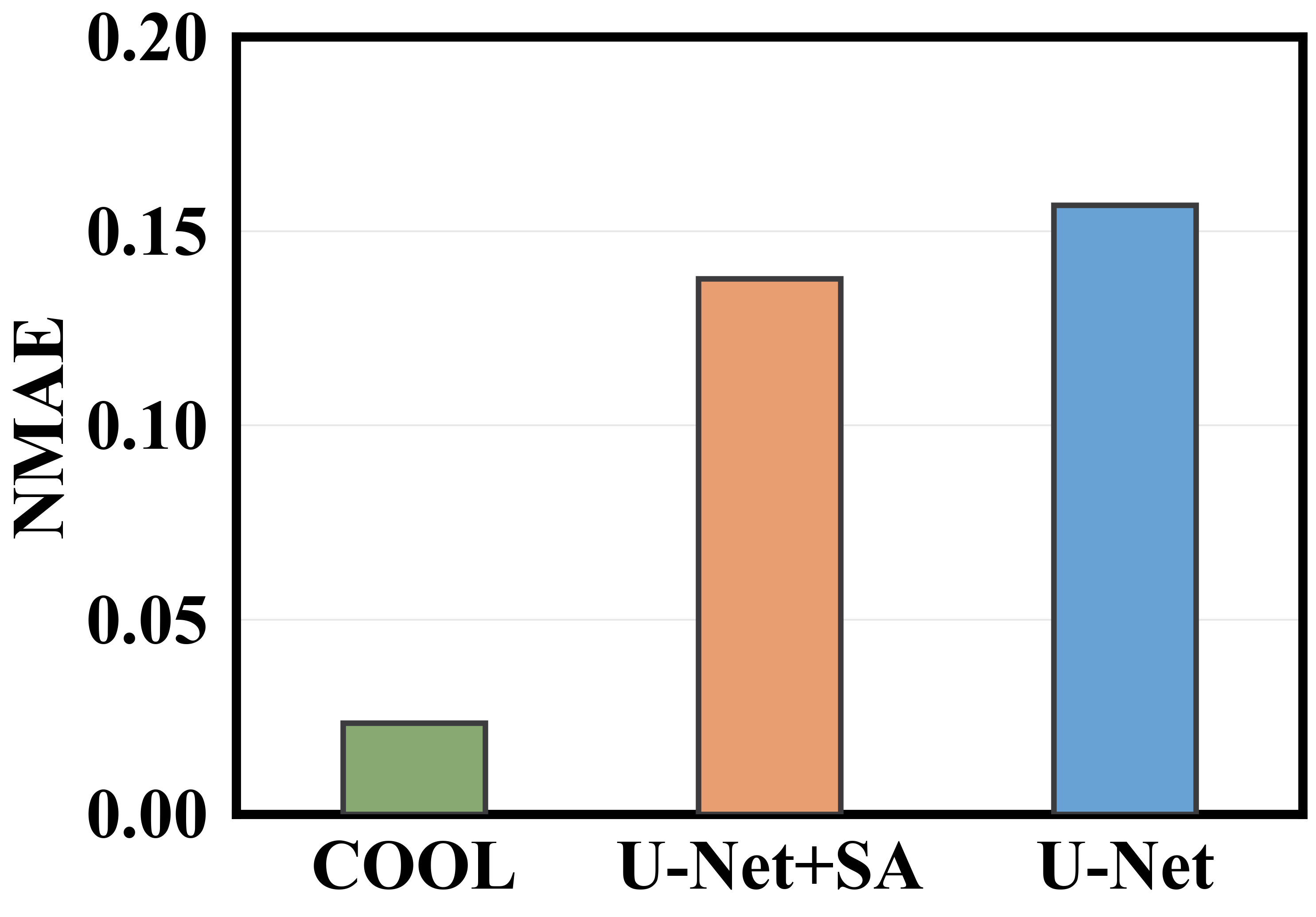}
  \caption{}
  \label{fig:r1}
\end{subfigure} 
\begin{subfigure}{0.49\linewidth}
  \centering
  \includegraphics[width=\linewidth]{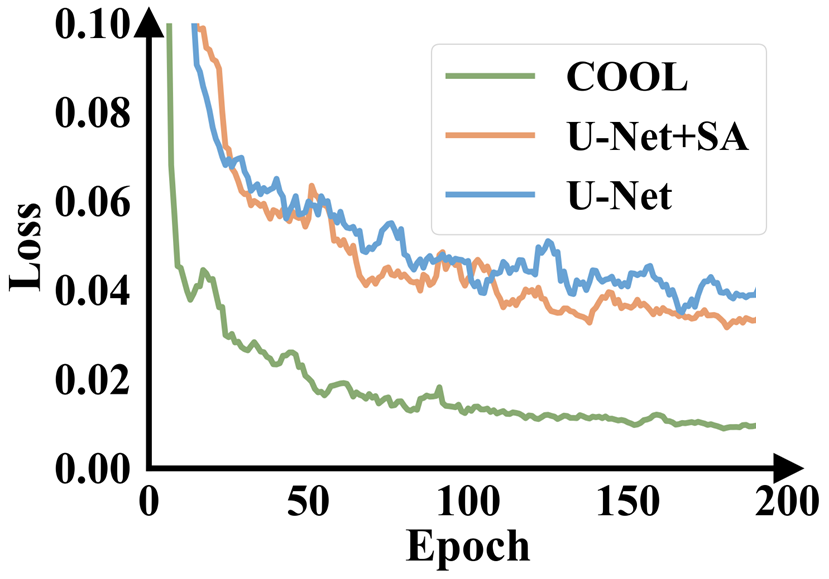}
  \caption{}
  \label{fig:r2}
\end{subfigure}
\vspace{-.2in}
\caption{Comparison among COOL (ours), U-Net, and U-Net+SA. 
    (a) In the evaluation stage, the testing results show that COOL consistently outperforms the two baselines, confirming that our method achieves better overall prediction accuracy. (b) During the training stage, COOL exhibits a faster and more stable convergence in training loss. }
\label{fig:loss}
\end{figure}

\begin{figure*}[!t]
\centering
\vspace{-.5in}
\begin{subfigure}{1\linewidth}
  \centering
  \includegraphics[width=\linewidth]{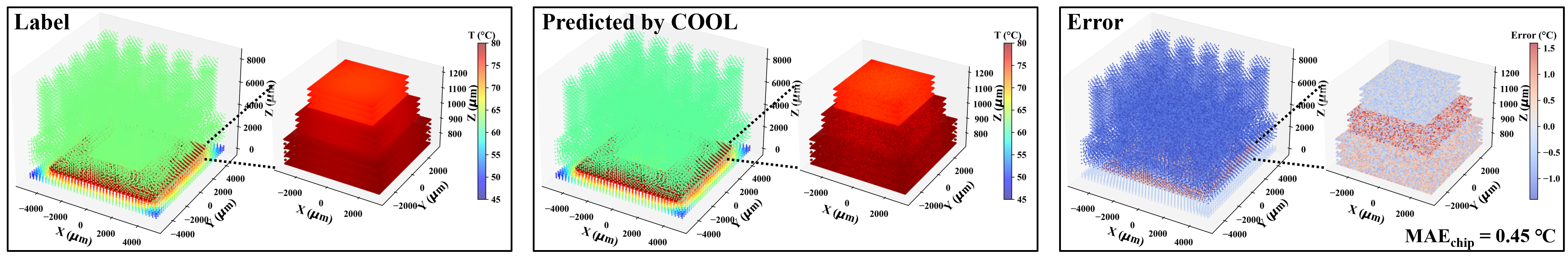}
  \vspace{-.2in}
  \caption{The first case study from our designs: $N_{\text{stacks}}=1$, $N_{\text{die}}=3$.}
  \label{fig:v1}
\end{subfigure} 
\\
\begin{subfigure}{1\linewidth}
  \centering
  \includegraphics[width=\linewidth]{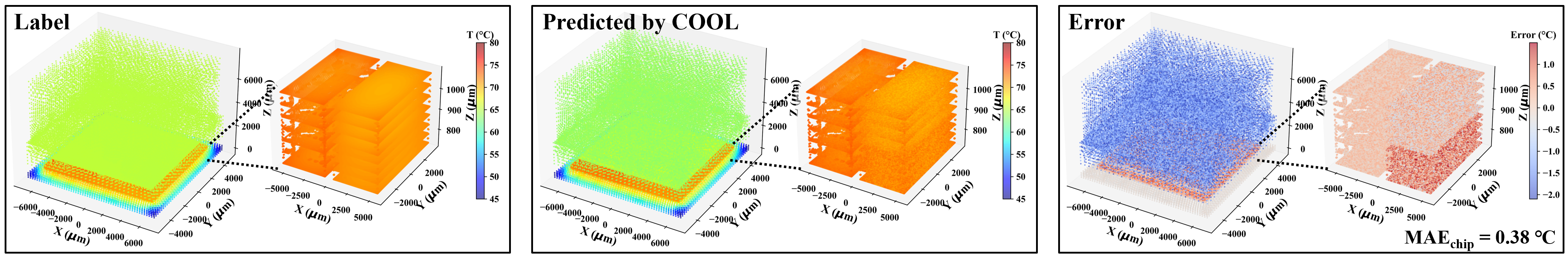}
  \vspace{-.2in}
  \caption{The second case study from our designs: $N_{\text{stacks}}=2$, $N_{\text{die}}=2$.}
  \label{fig:v2}
\end{subfigure}
\vspace{-.3 in}
\caption{Three-dimensional point cloud result visualization of temperature for specific designs. 
    }
\label{fig:result}
\vspace{-.15in}
\end{figure*}

\subsubsection{Error Metrics}






We evaluate prediction accuracy using the \textbf{mean absolute error (MAE)} and its range-normalized counterpart, the \textbf{normalized MAE (NMAE)}.  
MAE measures the average absolute difference between predicted and ground-truth temperatures, while NMAE provides a scale-independent error expressed as a percentage of the thermal range.  
They are defined as
\begin{equation}
\text{MAE} = \frac{1}{N}\sum_{i=1}^{N} |T_i - \hat{T}_i| ,\,\,\,
\text{NMAE} = \frac{\text{MAE}}{T_{\max} - T_{\min}}\times100\%
\end{equation}

where $\hat{T}_i$ and $T_i$ and are the predicted and label temperatures at point $i$, $N$ is the number of sampled points, and $T_{\max}$ and $T_{\min}$ denote the maximum and minimum temperatures of the design.

\subsection{Baseline}
We use commercial FEM tools to generate the temperature field as ground truth.
Since SAN-FNO~\cite{huang2025self} is originally designed for grid-based inputs and cannot directly handle point-cloud data, we construct comparable point-based baselines for comparison. Specifically, we implement a plain U-Net and an enhanced U-Net+SA (U-Net with Self-Attention) following similar design principles. These two models serve as point-based counterparts to SAN-FNO~\cite{huang2025self}, enabling a fair evaluation against our proposed COOL framework.

\subsection{Results}
In order to validate the effectiveness of COOL, we conducted experiments on heterogeneous IC designs under the experimental settings described in Section~\ref{sec:setup}. The aim was to assess both the predictive accuracy and computational efficiency of COOL.

Figure~\ref{fig:loss} presents a general comparison between COOL and baseline methods during both training and testing.  
It is evident from the figure that COOL exhibits a consistently lower loss trajectory throughout the process, suggesting more stable learning behavior.

Figure~\ref{fig:r1} reports testing results on the NMAE metric at \textit{epoch = 200}. 
COOL consistently outperforms the baselines, achieving an NMAE of approximately 2.4\%, while U-Net and U-Net+SA reach 15.6\% and 13.7\%, respectively. 
This substantial margin highlights COOL's superior predictive capability and confirms that it can accurately capture the underlying thermal patterns from point-cloud input data. Importantly, this performance gap suggests that traditional point-based and U-Net-based approaches struggle to generalize across the full spatial variability of heterogeneous IC designs, whereas COOL effectively models these complexities.

Figure~\ref{fig:r2} illustrates the training dynamics of COOL compared to point-based baselines, U-Net and U-Net+SA, over the first 200 epochs. COOL's loss decreases rapidly and approaches convergence around epoch 160, whereas both U-Net and U-Net+SA continue to decline noticeably, indicating slower convergence. 
This rapid convergence of COOL not only reduces the required training time but also suggests improved optimization stability, potentially leading to more robust performance across different IC designs.

Figure~\ref{fig:result} provides a visualization of COOL's prediction performance on two heterogeneous IC designs. The figure includes both full-chip temperature prediction results and detailed temperature distributions over the primary heat sources (chip regions). Figure~\ref{fig:v1} corresponds to a fully three-dimensional stacked IC, while Figure~\ref{fig:v2} shows a 3.5D configuration with a separate chip module. COOL accurately captures spatial thermal gradients and temperature patterns in both scenarios, achieving particularly high fidelity in active chip regions with MAEs of approximately $0.45^{\circ}\mathrm{C}$ and $0.38^{\circ}\mathrm{C}$ for the 3D and 3.5D designs, respectively.

These visual results further confirm that COOL not only achieves low global error but also delivers precise temperature estimation in critical regions, reinforcing its effectiveness for heterogeneous ICs.

We also provide a rough estimate of runtime. Using a commercial FEM solver as a reference, predicting each design typically requires 189 seconds, excluding manual interventions such as mesh refinement or parameter tuning. COOL completes prediction in 12 seconds per design without any manual intervention, resulting in a speedup greater than 15.7$\times$.
The combination of rapid convergence, high predictive accuracy, and low computational cost demonstrates that COOL provides a practical solution for real-time thermal analysis and design verification in heterogeneous ICs.

\section{Conclusion}


In this work, we propose COOL, a cooling-aware point transformer for thermal prediction in advanced 3D/3.5D IC packaging. By representing the thermal domain as an annotated point cloud, COOL explicitly models cooling structures as dynamic design elements. The proposed PI-BC scheme enforces physical consistency at material interfaces and cooling boundaries. Extensive experiments show COOL achieves 2.4\% NMAE, substantially outperforming point-based baselines while providing 15.7× speedup over FEM solvers. These results confirm COOL's practicality for thermal-aware co-design in next-generation IC packaging.

\section*{Acknowledgement}
This work is supported by Hong Kong Research Grants Council (RGC) CRF-YCRG C6003-24Y, GRF 16216825, and T46-415/25-R. It was partially conducted by ACCESS – AI Chip Center for Emerging Smart Systems, supported by the InnoHK initiative of the Innovation and Technology Commission of the Hong Kong Special Administrative Region Government.

\bibliographystyle{ACM-Reference-Format}
\bibliography{references_1, references_2}

@article{Bronstein_2017,
   title={Geometric Deep Learning: Going beyond Euclidean data},
   volume={34},
   ISSN={1558-0792},
   url={http://dx.doi.org/10.1109/MSP.2017.2693418},
   DOI={10.1109/msp.2017.2693418},
   number={4},
   journal={IEEE Signal Processing Magazine},
   publisher={Institute of Electrical and Electronics Engineers (IEEE)},
   author={Bronstein, Michael M. and Bruna, Joan and LeCun, Yann and Szlam, Arthur and Vandergheynst, Pierre},
   year={2017},
   month=jul, pages={18–42} }

@misc{qi2017pointnetdeephierarchicalfeature,
      title={PointNet++: Deep Hierarchical Feature Learning on Point Sets in a Metric Space}, 
      author={Charles R. Qi and Li Yi and Hao Su and Leonidas J. Guibas},
      year={2017},
      eprint={1706.02413},
      archivePrefix={arXiv},
      primaryClass={cs.CV},
      url={https://arxiv.org/abs/1706.02413}, 
}

@misc{pfaff2021learningmeshbasedsimulationgraph,
      title={Learning Mesh-Based Simulation with Graph Networks}, 
      author={Tobias Pfaff and Meire Fortunato and Alvaro Sanchez-Gonzalez and Peter W. Battaglia},
      year={2021},
      eprint={2010.03409},
      archivePrefix={arXiv},
      primaryClass={cs.LG},
      url={https://arxiv.org/abs/2010.03409}, 
}

@inproceedings{chen2023fast,
  title={Fast full-chip parametric thermal analysis based on enhanced physics enforced neural networks},
  author={Chen, Liang and Lu, Jincong and Jin, Wentian and Tan, Sheldon X-D},
  booktitle={2023 IEEE/ACM International Conference on Computer Aided Design (ICCAD)},
  pages={1--8},
  year={2023},
  organization={IEEE}
}

@inproceedings{feng2022chiplet,
  title={Chiplet actuary: A quantitative cost model and multi-chiplet architecture exploration},
  author={Feng, Yinxiao and Ma, Kaisheng},
  booktitle={Proceedings of the 59th ACM/IEEE Design Automation Conference},
  pages={121--126},
  year={2022}
}

@article{chen2020fast,
  title={A fast semi-analytic approach for combined electromigration and thermomigration analysis for general multisegment interconnects},
  author={Chen, Liang and Tan, Sheldon X-D and Sun, Zeyu and Peng, Shaoyi and Tang, Min and Mao, Junfa},
  journal={IEEE Transactions on Computer-Aided Design of Integrated Circuits and Systems},
  volume={40},
  number={2},
  pages={350--363},
  year={2020},
  publisher={IEEE}
}

@article{pak2014electromigration,
  title={Electromigration study for multiscale power/ground vias in TSV-based 3-D ICs},
  author={Pak, Jiwoo and Lim, Sung Kyu and Pan, David Z},
  journal={IEEE Transactions on Computer-Aided Design of Integrated Circuits and Systems},
  volume={33},
  number={12},
  pages={1873--1885},
  year={2014},
  publisher={IEEE}
}

@inproceedings{shao2019simba,
  title={Simba: Scaling deep-learning inference with multi-chip-module-based architecture},
  author={Shao, Yakun Sophia and Clemons, Jason and Venkatesan, Rangharajan and Zimmer, Brian and Fojtik, Matthew and Jiang, Nan and Keller, Ben and Klinefelter, Alicia and Pinckney, Nathaniel and Raina, Priyanka and others},
  booktitle={Proceedings of the 52nd annual IEEE/ACM international symposium on microarchitecture},
  pages={14--27},
  year={2019}
}

@inproceedings{wang2024exploiting,
  title={Exploiting 2.5 d/3d heterogeneous integration for ai computing},
  author={Wang, Zhenyu and Sun, Jingbo and Goksoy, Alper and Mandal, Sumit K and Liu, Yaotian and Seo, Jae-Sun and Chakrabarti, Chaitali and Ogras, Umit Y and Chhabria, Vidya and Zhang, Jeff and others},
  booktitle={2024 29th Asia and South Pacific Design Automation Conference (ASP-DAC)},
  pages={758--764},
  year={2024},
  organization={IEEE}
}

@inproceedings{chen2025neuralmesh,
  title={NeuralMesh: Neural Network For FEM Mesh Generation in 2.5 D/3D Chiplet Thermal Simulation},
  author={Chen, Pengju and Niu, Dan and Zhang, Dekang and Wang, Wenhao and Xie, Depeng and Jin, Zhou and Xing, Wei W and He, Lei},
  booktitle={2025 62nd ACM/IEEE Design Automation Conference (DAC)},
  pages={1--7},
  year={2025},
  organization={IEEE}
}

@inproceedings{Sun2025ChipletEMP2,
  title={ChipletEM: Physics-Based 2.5D and 3D Chiplet Heterogeneous Integration Electromigration Signoff Tool Using Coupled Stress and Thermal Simulation},
  author={Zeyu Sun and Weijie Tong and Xiaoning Ma and He Cao and Jianyun Liu and Zhiqiang Li and Qinzhi Xu},
  booktitle={2025 62nd ACM/IEEE Design Automation Conference (DAC)},
  pages={1--7},
  year={2025},
  organization={IEEE}
}

@inproceedings{huang2025self,
  title={Self-Attention to Operator Learning-based 3D-IC Thermal Simulation},
  author={Huang, Zhen and Wang, Hong and Yang, Wenkai and Tang, Muxi and Xie, Depeng and Lin, Ting-Jung and Zhang, Yu and Xing, Wei W and He, Lei},
  booktitle={2025 62nd ACM/IEEE Design Automation Conference (DAC)},
  pages={1--7},
  year={2025},
  organization={IEEE}
}

@inproceedings{zhou2025asrr,
  title={ASRR-PINN: Adaptive Sub-Regional Random Resampling-Based PINN for Thermal Analysis of 3D-ICs},
  author={Zhou, Zijian and Tang, Min and Chen, Liang},
  booktitle={2025 62nd ACM/IEEE Design Automation Conference (DAC)},
  pages={1--7},
  year={2025},
  organization={IEEE}
}

@article{wang2025hisim,
  title={HISIM: Analytical Performance Modeling and Design Space Exploration of 2.5 D/3D Integration for AI Computing},
  author={Wang, Zhenyu and Nalla, Pragnya Sudershan and Sun, Jingbo and Goksoy, A Alper and Mandal, Sumit K and Seo, Jae-sun and Chhabria, Vidya A and Zhang, Jeff and Chakrabarti, Chaitali and Ogras, Umit Y and others},
  journal={IEEE Transactions on Computer-Aided Design of Integrated Circuits and Systems},
  year={2025},
  publisher={IEEE}
}

@inproceedings{chen2022fast,
  title={Fast thermal analysis for chiplet design based on graph convolution networks},
  author={Chen, Liang and Jin, Wentian and Tan, Sheldon X-D},
  booktitle={2022 27th Asia and South Pacific Design Automation Conference (ASP-DAC)},
  pages={485--492},
  year={2022},
  organization={IEEE}
}

@article{yu2025deepoheat,
  title={DeepOHeat-v1: Efficient Operator Learning for Fast and Trustworthy Thermal Simulation and Optimization in 3D-IC Design},
  author={Yu, Xinling and Liu, Ziyue and Li, Hai and Li, Yixing and Ai, Xin and Zeng, Zhiyu and Young, Ian and Zhang, Zheng},
  journal={arXiv preprint arXiv:2504.03955},
  year={2025}
}

@inproceedings{liu2023deepoheat,
  title={DeepOHeat: operator learning-based ultra-fast thermal simulation in 3D-IC design},
  author={Liu, Ziyue and Li, Yixing and Hu, Jing and Yu, Xinling and Shiau, Shinyu and Ai, Xin and Zeng, Zhiyu and Zhang, Zheng},
  booktitle={2023 60th ACM/IEEE Design Automation Conference (DAC)},
  pages={1--6},
  year={2023},
  organization={IEEE}
}

@inproceedings{wu2024point,
  title={Point transformer v3: Simpler faster stronger},
  author={Wu, Xiaoyang and Jiang, Li and Wang, Peng-Shuai and Liu, Zhijian and Liu, Xihui and Qiao, Yu and Ouyang, Wanli and He, Tong and Zhao, Hengshuang},
  booktitle={Proceedings of the IEEE/CVF conference on computer vision and pattern recognition},
  pages={4840--4851},
  year={2024}
}

@inproceedings{zhao2021point,
  title={Point transformer},
  author={Zhao, Hengshuang and Jiang, Li and Jia, Jiaya and Torr, Philip HS and Koltun, Vladlen},
  booktitle={Proceedings of the IEEE/CVF international conference on computer vision},
  pages={16259--16268},
  year={2021}
}
\end{document}